\documentclass[twocolumn,showpacs,prl,floatfix,letter]{revtex4-1}
\ifx\pdfoutput\undefined
\usepackage{graphicx}
\else
\usepackage{epstopdf}
\usepackage{epsfig}
\usepackage{amssymb,amsmath,stmaryrd,tabularx}
\usepackage{wrapfig}
\usepackage{bm}
\fi

\usepackage[center]{subfigure}

\begin{document}
\title{Rare fluctuations and large-scale circulation cessations in turbulent convection}
\author{Michael Assaf$^1$, Luiza Angheluta$^{1,2}$ and Nigel Goldenfeld$^1$}
\affiliation{$^1$Department of Physics, University of Illinois at
Urbana-Champaign, Loomis Laboratory of Physics, 1110 West Green
Street, Urbana, Illinois, 61801-3080\\
$^2$Physics of Geological Processes, Department of Physics, University of Oslo, Norway}



\begin{abstract}
In turbulent Rayleigh-Benard convection, a large-scale circulation
(LSC) develops in a nearly vertical plane, and is maintained by rising
and falling plumes detaching from the unstable thermal boundary layers.
Rare but large fluctuations in the LSC amplitude can lead to extinction
of the LSC (a cessation event), followed by the re-emergence of another
LSC with a different (random) azimuthal orientation.  We extend
previous models of the LSC dynamics to include momentum and thermal
diffusion in the azimuthal plane, and calculate the tails of the
probability distributions of both the amplitude and azimuthal angle.
Our analytical results are in very good agreement with experimental
data.
\end{abstract}
\maketitle

\def\Ra{\text{Ra}}
\def\Re{\text{Re}}

When a fluid is heated from below in the presence of a gravitational
field, the static state with thermal conduction can become unstable
towards a succession of instabilites, ultimately leading to turbulence
if the buoyancy-induced driving force is sufficiently greater than the
viscous drag and diffusion of heat.  This balance is quantified by the
Rayleigh number $\Ra = \alpha_{_0} g\Delta T L^3/\nu\kappa$, where $\alpha_{_0}$
is the isobaric thermal expansion coefficient, $g$ is the gravity
field, $\Delta T$ is the temperature gap between bottom and top layers,
$L$ is the height of the fluid container, $\kappa$ is the thermal
diffusivity and $\nu$ is the kinematic viscosity.  For large Ra,
thermal boundary layers become unstable by emitting hot (on the bottom)
or cold (on the top) plumes which, due to buoyancy, migrate upwards
(hot) or downwards (cold)~\cite{Krishnamurti81,SIGG94,AHLE09}. In
addition to their vertical motion, plumes drift along the top and
bottom boundaries in opposite directions, contributing to a large-scale
circulation (LSC) flowing in a nearly vertical plane, which spans the
diameter of the container. The horizontal velocity of the plumes
oscillates rapidly compared to the reorientation dynamics of the
large-scale circulation~\cite{Ahlers04,Xi09,Ahlers09}, which, in a
cylindrical geometry, undergoes both rotational diffusion and
orientational jumps following irregular cessation of the entire
flow~\cite{Ahlers04,Ahlers06}. Such laboratory experiments provide a
well-controlled setting in which to study the statistical properties of
cessation, reversal and reorientation events similar to those that
occur in many flows of practical significance, including
atmospheric~\cite{Vand00} and oceanic circulation~\cite{MARS99}, the
dynamo driving planetary magnetic fields~\cite{ROBE00}, and in the
cores of stars~\cite{MIES09}.

In order to interpret high quality data on the statistics of cessations
and azimuthal rotation, a nonlinear stochastic model that retains
physically relevant aspects of the Navier-Stokes equations was
developed and shown to reproduce many aspects of the statistics of the
azimuthal dynamics and the temperature fluctuations in the LSC
plane~\cite{Ahlers07, Ahlers08}. The stochastic variables in the model
are the amplitude of azimuthal temperature variations, $\delta$,
induced by the LSC and the azimuthal orientation angle, $\theta_0$, of
the nearly vertical LSC plane. Although the model predictions are in
good agreement with the experimental results for typical
fluctuations of the system~\cite{Ahlers08}, the model does not account
quantitatively for the rare large fluctuations responsible for the
cessation statistics and for the broad-tail probability distribution
function (PDF) of the azimuthal velocities.

The purpose of this Letter is to extend the stochastic model to capture
the tail of the PDFs of the temperature amplitudes and azimuthal
velocities.  We make three contributions here. Firstly, we show that
the equation for the amplitude $\delta$ needs to explicitly include a
constant term, known to scale as $\Ra^{5/4}$. Such a term was already
proposed in Ref.~\cite{Ahlers08} as arising from boundary layer thermal
diffusion, but its significance for the asymptotics of the PDF had not
been emphasized.  Secondly, we show that the description of the
azimuthal velocities needs to include viscous diffusion in the boundary
layer near the wall. Such a term is generally small compared to the
other terms in the equation of motion for $\dot\theta_0$, but becomes
the dominant contribution when the amplitude $\delta$ is small, as in a
cessation event.  Thirdly, we compute the PDFs for both $\delta$ and
$\dot\theta_0$, predicting respectively an exponential dependence at
small $\delta$ and a power law of $-4$ for the large angular velocity asymptotics
of $\dot\theta_0$.  A careful analysis of the experimental data is in
very good agreement with these predictions.

{\it Evolution equation for the LSC amplitude:-\/} For completeness, we
briefly summarize the derivation of the physical model for LSC
fluctuations, largely following Refs.~\cite{Ahlers07, Ahlers08}, but
with minor differences noted below.  The LSC amplitude evolution is
derived from the equation satisfied by the velocity component in the
LSC plane, $u_\phi$, where only the buoyancy and diffusion terms are
retained. Here, $\phi$ is the angle in the vertical circulation plane of
the LSC. The turbulent advection term is discarded
on the basis that the convection due to azimuthal motion is small
relative to the other terms and the self-advection is replaced by
random fluctuations. A spatial average in a direction perpendicular to
the main axis of the cylinder (radial average) is performed. The
buoyancy term acts everywhere in the LSC plane, hence the average keeps
the same form. On the other hand, momentum diffusion is assumed to
dominate only in the viscous boundary layer, so that the average of
this term gives a prefactor $\lambda/L$, where $\lambda$ is the viscous
boundary layer thickness. The viscous layer thickness can be estimated
on dimensional grounds as the length scale where the convective forces
balance out the diffusive forces, giving ${U^2}/{L} \simeq \nu
{U}/{\lambda^2}$, where $U(t)$ is the maximum speed just within the
viscous boundary layer, and thus is an estimate for the typical
turnover velocity of an eddy spanning the $LSC$-plane.  Hence
$\lambda(t)\sim \sqrt{\nu L/U}$ and the spatial average of the
diffusion term is estimated as $\langle\nu\nabla^2u_\phi\rangle \simeq
-\nu U/(L\lambda)\simeq -\nu^{1/2}U^{3/2}/L^{3/2}$. Furthermore, we
assume that the amplitude of azimuthal temperature variation, $\delta$,
is proportional to the large-scale typical velocity of thermal
convection rolls $U$, in the approximation that momentum acceleration
is due to buoyancy forces; this proportionality argument is different
from one used in Ref.~\cite{Ahlers08}, where buoyancy is balanced
against diffusion. Finally, a delta-correlated Gaussian stochastic forcing
$f_{\delta}(t)$ with amplitude $D_{\delta}$ is included to simulate
the effect of turbulent fluctuations.  As noted in
Ref.~\cite{Ahlers08}, the resulting equation incorrectly accounts for
the small $\delta$ behavior, where the thermal boundary layer cannot be
neglected.  Thermal diffusion leads to a constant driving term
$\dot{\delta}=A$ empirically~\cite{Ahlers06} found to scale as
$\Ra^{5/4}$.  This has the effect of driving the system back to the
vicinity of $\delta=\delta_0$, where the mean LSC amplitude is denoted
by $\delta_0 \approx {\Delta T\sigma}\Re^{3/2}/{\Ra}$ and
$\sigma=\nu/\kappa$ is the Prandtl number. By rescaling time $t\to
t/\tau_{\delta}$, where $\tau_\delta \approx L^2/(\nu \Re^{1/2})$ is the
typical turnover time, and defining a dimensionless amplitude
$\xi=\delta/\delta_0$, we arrive at the following Langevin equation for
the LSC fluctuations $\xi$
\begin{equation}\label{ratexi}
\dot{\xi}=\tilde{A}+\alpha\xi-\beta\xi^{3/2}+\tilde{f}_{\xi}(t),
\end{equation}
with $\tilde{A}=A\tau_{\delta}/\delta_0$.  Here the rescaled
(dimensionless) diffusion coefficient is $\tilde{D}_{\delta}\equiv
D_{\delta}\tau_{\delta}/\delta_0^2$, representing the amplitude of the
scaled noise $\tilde{f}_{\xi}(t)$.  We have included numerical
prefactors $\alpha,\beta={\cal O}(1)$ to account for the geometric
coefficients from the spatial volume averaging procedure. These
constants will be determined below by demanding that the maximum and
the width of the PDF are consistent with experimental results.

{\it Evolution of the azimuthal velocity:-\/} The equation for the
horizontal motion is obtained from the Navier-Stokes equation for the
azimuthal velocity, $u_\theta\simeq L\dot\theta$, by retaining the
advection and momentum diffusion terms.
Previously~\cite{Ahlers07,Ahlers08}, the viscous drag term was
neglected on the basis that it is typically small. This approximation
is valid in the regime of a well-defined LSC, but breaks down near
cessations, since the momentum transport from the LSC also becomes very
small. The viscous drag is dominant in the viscous boundary layer, so
that a spatial average along an arbitrary direction in the horizontal
plane gives
$\langle\nu\nabla^2\dot\theta_0\rangle\sim-\nu\dot\theta_0/(L\lambda_\theta)$.
The viscous boundary layer thickness $\lambda_\theta$ is estimated from
balancing the advection force with the momentum diffusion force $U
\dot\theta_0/L \sim \nu \dot\theta_0/\lambda_\theta^2$; together with
the proportionality $U/\tau_\delta\sim \alpha g \delta$, we find
that $\lambda_\theta \sim \sqrt{\nu L\delta_0/(U_0\delta)}$. In
addition to these deterministic forces, the self-advection term is
mimicked by a delta-correlated Gaussian noise $f_{\dot\theta}(t)$ with amplitude
$D_{\dot\theta}$. Rescaling time by the typical time
$\tau_\theta \approx L^2/(\nu \Re)$ for crossing a boundary layer of thickness $\lambda_\theta$ and $\delta$ by $\delta_0$, the equation of motion for the azimuthal fluctuations is
\begin{equation}\label{thetaeq}
\ddot \theta_0  = -\left(\alpha_1\xi +\beta_1\frac{\tau_\theta}{\tau_\delta}\sqrt{\xi}\right)\dot\theta_0+\tilde f_{\dot\theta}(t),
\end{equation}
where the rescaled diffusion coefficient is
$\tilde{D}_{\dot{\theta}}=D_{\dot{\theta}}\tau_\theta$, and
$\alpha_1,\beta_1={\cal O}(1)$ account for geometrical factors due to
volume averaging. From the definition of the timescales,
$\tau_\theta/\tau_\delta = \Re^{-1/2}\ll 1$ and hence the viscous drag
term becomes important when $\xi \ll (\beta_1/\alpha_1)^2\Re^{-1}$,
i.e. near cessations.

{\it Probability distribution for $\delta$:-\/} Since the Langevin
equation for $\xi$ is decoupled from that of $\dot{\theta}$, we first
analyze Eq.~(\ref{ratexi}) separately. In order to obtain the
stationary PDF $P(\xi)$ at long times, we use the equivalent
Fokker-Planck equation of Eq.~(\ref{ratexi}). It reads~\cite{Gardiner}
\begin{equation}\label{FP}
\frac{\partial P(\xi,t)}{\partial t}=-\frac{\partial}{\partial
\xi}\left[\left(\tilde{A}\!+\!\alpha\xi\!-\!\beta\xi^{3/2}\right)P(\xi,t)\right]+\frac{\tilde{D}_{\delta}}{2}\frac{\partial^2
P(\xi,t)}{\partial \xi^2}.
\end{equation}
The stationary solution of this equation is
\begin{equation}\label{pdf0}
P(\xi)=C \exp[-2V(\xi)/\tilde{D}_{\delta}],
\end{equation}
with
\begin{equation}
V(\xi)=-\tilde{A}\xi-\alpha\frac{\xi^2}{2}+\beta\frac{2}{5}\xi^{5/2}.
\label{eq_potential}
\end{equation}
Note that Eq.~(\ref{pdf0}) predicts that $\log P(\xi\ll 1) \propto \xi$
as observed in experiment. Denoting the logarithmic derivative of the
experimental PDF at small $\xi$ by $B$, using~(\ref{pdf0}) we find that
$\tilde{A}=B\tilde{D}_{\delta}/2$. Here $B$ and $\tilde{D}_{\delta}$
are the tuning parameters of the theory and will be extracted from
experimental data.
\begin{figure}[t]
\includegraphics[width=\columnwidth]{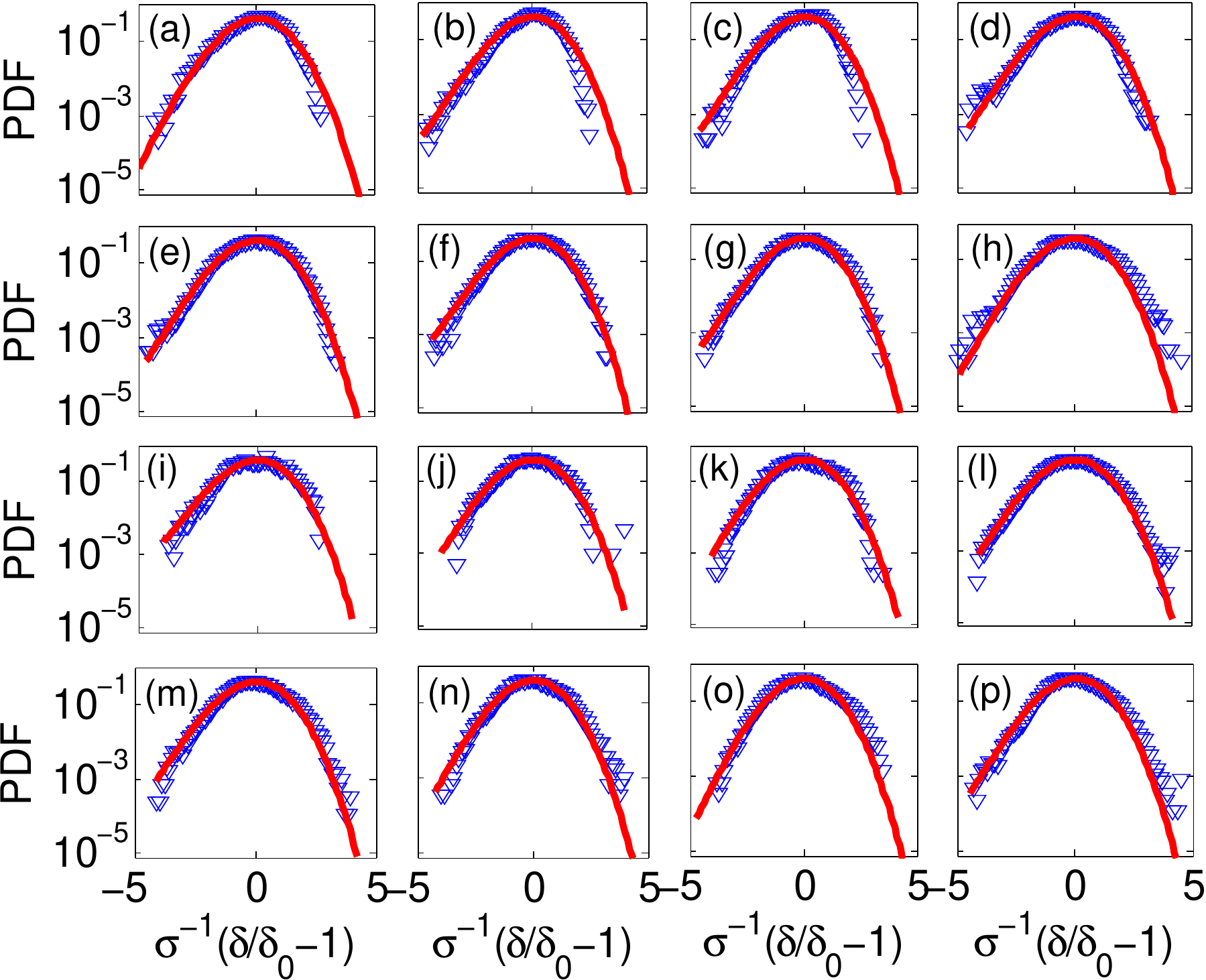}
\caption{(Color online) Theoretical (solid line) and
experimental (triangles) PDFs for the normalized amplitude $\xi$
versus $(\xi-1)/\sigma$ ($\sigma=\sqrt{\tilde{D}_{\delta}}$), for different $\Ra$ numbers, with (a)-(h) for
the medium sample and (i)-(p) for the large sample. The $\Ra$ numbers
are: $3.78\cdot 10^8$ (a), $8.16\cdot 10^8$ (b), $1.1\cdot 10^9$ (c),
$2.3\cdot 10^9$ (d), $4.5\cdot 10^9$ (e), $7.9\cdot 10^9$ (f),
$1.02\cdot 10^{10}$ (g), $1.51\cdot 10^{10}$ (h), $4.75\cdot 10^9$ (i),
$7.16\cdot 10^9$ (j), $1.22\cdot 10^{10}$ (k), $2.43\cdot 10^{10}$ (l),
$4.71\cdot 10^{10}$ (m), $5.68\cdot 10^{10}$ (n), $7.51\cdot 10^{10}$
(o) and $1.04\cdot 10^{11}$ (p). In each subfigure the parameters
$\tilde{D}_{\delta}$ and $B$ were computed by fitting the left tail of
the experimental PDF to Eq.~(\ref{lefttail}).} \label{pdffit}
\end{figure}

We now determine the constants $\alpha$ and $\beta$ by requiring that
the PDF has a maximum at $\xi=1$ and width equal to
$\sqrt{\tilde{D}_{\delta}}$, and fix the constant $C$ by normalizing
$P(\xi)$ in its Gaussian regime close to $\xi=1$. Expanding
$P(\xi)$~(\ref{pdf0}) in the vicinity of $\xi=1$ up to second order, we
find that $\tilde{A}+\alpha-\beta=0$ for the maximum to be at $\xi=1$,
and $(3/2)\beta-\alpha=1/2$ for the variance to be
$\tilde{D}_{\delta}$. This yields $\alpha=1-3\tilde{A}$ and
$\beta=1-2\tilde{A}$. With $\tilde{A}=B\tilde{D}_{\delta}/2$, the final
normalized result for the PDF reads as
\begin{eqnarray}\label{pdf2}
P(\xi)&=&\frac{1}{\sqrt{2\pi \tilde{D}_{\delta}}}e^{-3B/10-1/(5\tilde{D}_{\delta})}\nonumber\\
&\times& e^{B\xi+\tilde{D}_{\delta}^{-1}\left[\left(1-3B\tilde{D}_{\delta}/2\right)\xi^2-(4/5)\left(1-B\tilde{D}_{\delta}\right)\xi^{5/2}\right]}.
\end{eqnarray}
The only free parameters in this result are $B$ and
$\tilde{D}_{\delta}$. These are estimated from the $\xi\ll 1$
asymptote of~(\ref{pdf2}):
\begin{equation}\label{lefttail}
P(\xi\ll 1)\simeq (2\pi \tilde{D}_{\delta})^{-1/2}e^{-3B/10-1/(5\tilde{D}_{\delta})}\,e^{B\xi}.
\end{equation}
By fitting the logarithm of the experimental PDFs to a line
from the logarithm of Eq.~(\ref{lefttail}), we extract $B$ and
$\tilde{D}_{\delta}$ for each experimental data set. In
Fig.~\ref{pdffit} we show comparisons of experimental results and
PDF~(\ref{pdf2}) using the parameters $B$ and $\tilde{D}_{\delta}$
extracted from experimental results; very good agreement is
evident for a wide range of $\Ra$ numbers. In this figure and henceforth the medium and large samples refer to cylindrical containers with heights $24.76$cm and $50.61$cm and aspect ratio of $\sim 1$~\cite{Ahlers08}.
\begin{figure}[t]
\includegraphics[width=\columnwidth]{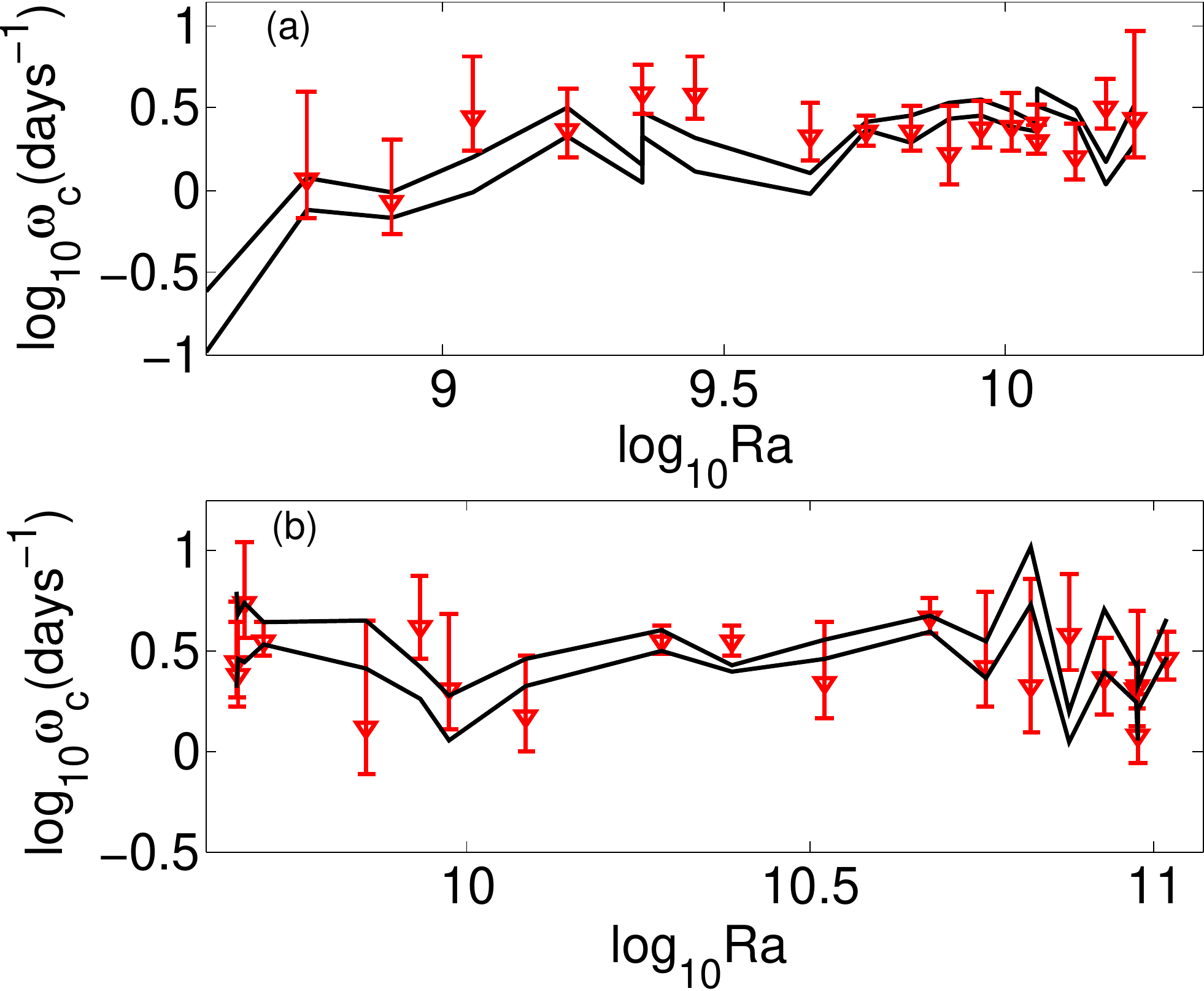}
\caption{(Color online) Cessation frequency (per day) as a
function of $\Ra$ for the medium (a) and large (b) samples. Experimental
results (triangles)~\cite{Ahlers08} are compared to
theoretical prediction~(\ref{cesstime1}). The latter are bound within the two solid lines. The experimental cessation
was defined to occur when $\delta/\delta_0<\xi_{min}$, that is, we have averaged
over the time intervals between events where the system undergone
cessation with $\xi<\xi_{min}$, see Eq.~(\ref{cesstime1}). Here, $\xi_{min}=0.15$ for the medium sample, and $\xi_{min}=0.2$ for the large sample. Similar results were obtained for thresholds of $\xi_{min}$ between 0.15 and 0.3.}  \label{cessfreq}
\end{figure}

Having calculated the complete PDF of the LSC amplitudes, we now extract the
cessation frequency. The latter can be found by analyzing the following
first-passage problem: starting from the vicinity of the fixed point
$\xi=1$ what is the mean time it takes to reach the vicinity of
$\xi=\xi_0\ll 1$, where $\xi_0\ll 1$ is the amplitude which defines the
experimental cessation threshold?  Writing down the backward
Fokker-Planck equation~\cite{Gardiner}, the mean time to cessation
(MTC) is given by
\begin{equation}
T(\xi,\xi_0)\!=\!2\int_{\xi_0}^{\xi}\frac{dy}{\psi(y)}\int_y^{\infty}\frac{\psi(z)}{\tilde{D}_{\delta}}dz\,;\;\;\psi(z)\!=\!e^{-\frac{2[V(z)-V(\xi_0)]}{\tilde{D}_{\delta}}},
\end{equation}
where $\xi\simeq 1$ is the effective initial condition, and the
potential satisfies~(\ref{eq_potential}) with
$\tilde{A}=B\tilde{D}_{\delta}/2$. Using the smallness of $\tilde{D}_{\delta}$ (typically ranging between $10^{-2}$--$10^{-1}$), we can evaluate the inner integral by using the saddle point approximation. By doing so, we arrive at a
result independent of $y$, which permits the evaluation of the outer
integral using a Taylor expansion of the integrand about
$y=\xi_0$~\cite{Assaf06}. This procedure leads to the final
result for the MTC $T_c(\xi_0)$ to reach a point $\xi_0\ll 1$ (see
also~\cite{Ahlers08}):
\begin{equation}\label{cesstime}
T_c (\xi_0)\simeq
\frac{\tau_{\delta}\tilde{D}_{\delta}}{|V'(\xi_0)|}\sqrt{\frac{2\pi}{\tilde{D}_{\delta}}}\,e^{2\tilde{D}_{\delta}^{-1}[V(\xi_0)-V(1)]},
\end{equation}
where we have multiplied the result by $\tau_{\delta}$ to present the time
in physical units, and used the fact that $V''(1)=1/2$. Given a threshold for cessation $\xi_{min}$ as is
done experimentally, in order to mimic the binning procedure of the
experimental data, we have to average over $\xi_0$ in
Eq.~(\ref{cesstime}) from $0$ to $\xi_{min}$, which yields
\begin{equation}\label{cesstime1}
\omega_c^{-1}\simeq \frac{1}{\xi_{min}}\int_0^{\xi_{min}} d\xi_0 \, T_c(\xi_0).
\end{equation}
To obtain a theoretical prediction for $\omega_c$ as a function of
$\Ra$, we use the extracted values of $B$, $\tilde{D}_{\delta}$ and $\tau_{\delta}$ from the experimental
data~\cite{Ahlers08}. By doing so, we can plot the theoretical prediction for $\omega_c$ as a function of $\Ra$, as shown in Fig.~\ref{cessfreq}. Here, the theoretical
predictions~(\ref{cesstime1}) agree well with the experimental data~\cite{Ahlers08}, for both the medium and large samples. The error
bars in the experimental results originate from the binning method,
while the errors in the theoretical curves come from the uncertainties
in the extracted values of $B$ and $\tilde{D}_{\delta}$.

\begin{figure}[t]
\includegraphics[width=0.8\columnwidth]{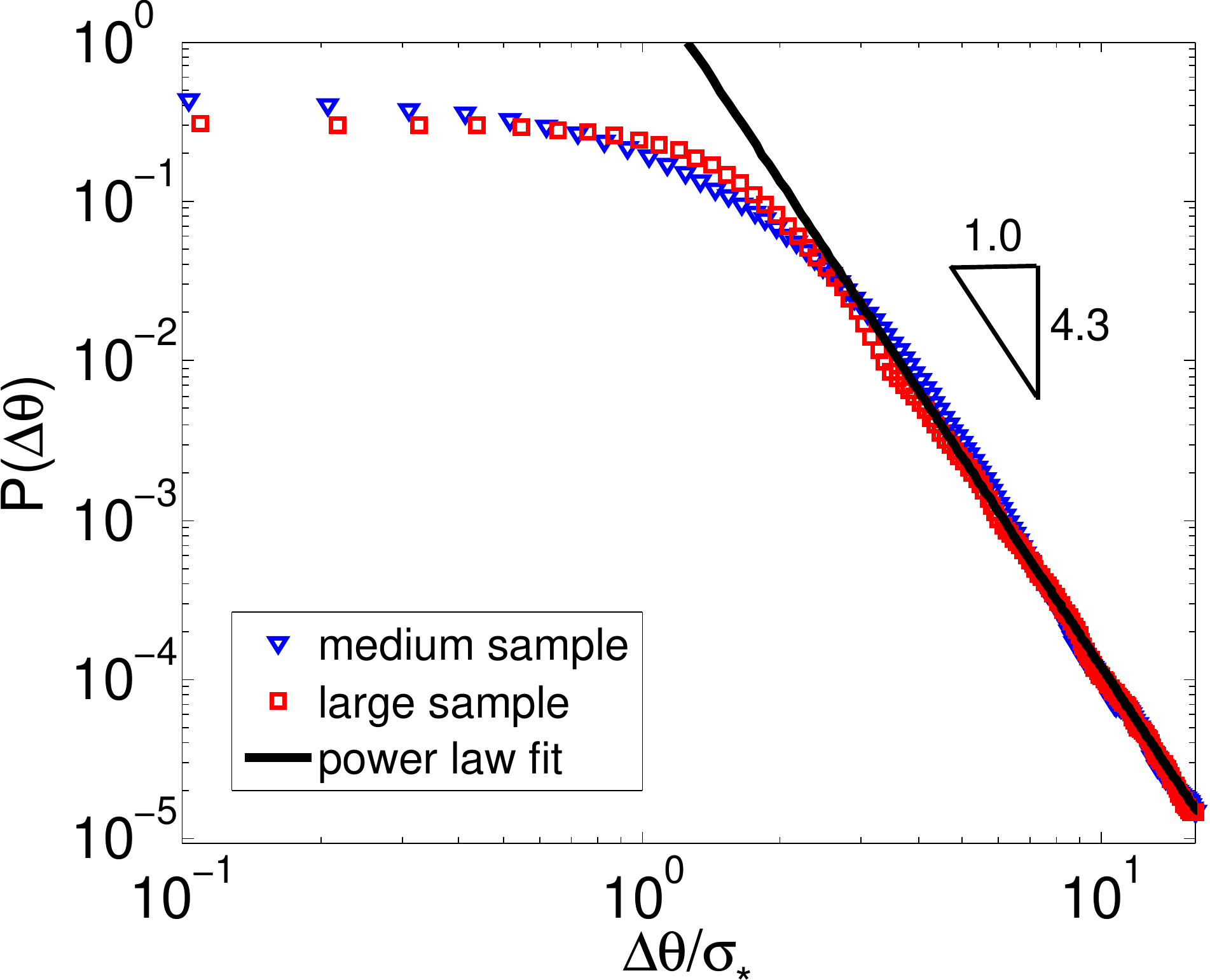}
\caption{(Color online) PDF $P(\Delta\theta)$ averaged over all
experimental PDFs with $\Ra$ numbers ranging from $10^9$-$10^{11}$ as a function of $\Delta\theta/\sigma_*$;
$\sigma_*$ is a rescaling factor so that the medium  (triangles)
 and large (squares) sample PDFs coincide at their right
tail. The experimental data shown here was adaptively binned using the data threshold method~\cite{Adami} with threshold $0.001\max [P(\Delta\theta)]$. Using the least-square method in the regime $4\lesssim \Delta\theta/\sigma_*\lesssim 20$, the medium and large  samples were found to scale with a power law of $-4.34\pm 0.02$ and $-4.27\pm 0.02$, respectively, closely fitting the theoretical prediction of Eq. (\ref{eq_thetamin4}). The solid line is a power law of $-4.3$.}
\label{pdftheta}
\end{figure}

{\it Probability distribution for $\dot\theta$:-\/} Now we
turn to the calculation of $P(\dot{\theta})$. As the Langevin
equation for $\dot{\theta}$ [Eq.~(\ref{thetaeq})] depends on $\xi$, for a
given $\xi$ we can first determine the steady state conditional
PDF $P(\dot\theta_0|\xi)$. Using~(\ref{thetaeq}), we find
\begin{equation}
P(\dot\theta_0|\xi) = \frac{1}{\sqrt{2\pi\tilde D_{\dot{\theta}}}}e^{-(\alpha_1\xi+\beta_1\sqrt{\xi/\Re})\dot\theta_0^2/\tilde D_{\dot{\theta}}}.
\end{equation}
Given the PDF $P(\xi)$ from Eq.~(\ref{pdf2}), we then determine the
complete PDF $P(\dot\theta_0)$ of the azimuthal velocity
by the following relation
\begin{equation}\label{eq:Ptheta}
P(\dot\theta_0) = \int_0^\infty d\xi P(\dot\theta_0|\xi) P(\xi),
\end{equation}
which is valid when the relaxation timescale of $\dot{\theta}$ is much
faster than that of $\delta$, namely $\tau_\delta\gg\tau_\theta$. That
is, Eq.~(\ref{eq:Ptheta}) holds when the conditional PDF
$P(\dot\theta_0|\xi)$ equilibrates much faster than the typical
timescale of change of $\xi$.

Integral~(\ref{eq:Ptheta}) can be evaluated in the Gaussian regime of
the PDF, where $\xi\simeq 1$, which yields the statistics of
reorientations due to rotations of the LSC plane. In this case, to
leading order one can simply put $\xi=1$ in Eq.~(\ref{thetaeq}) which
gives $P(\dot\theta_0)\sim
e^{-\alpha_1\dot\theta_0^2/\tilde{D}_{\dot{\theta}}}$. By comparing it
with the experiments, $\alpha_1=1$ in agreement with
Ref.~\cite{Ahlers08}.

The cessation events correspond to the right hand tail of
the PDF~(\ref{eq:Ptheta}). Indeed, when the system undergoes cessation
and $\xi\ll 1$, the integrand is dominated by
$e^{-\beta_1\sqrt{\xi/{\Re}}\dot\theta_0^2/\tilde D_{\dot{\theta}}}$. Therefore,
the right hand tail of the PDF given by Eq.~(\ref{eq:Ptheta}) satisfies
\begin{equation}
P(\dot\theta_0) \sim \dot\theta_0^{-4}.
\label{eq_thetamin4}
\end{equation}
This power-law prediction for the tail of $P(\dot\theta_0)$ is
consistent with the earliest analysis of the experimental
data~\cite{Ahlers06,Ahlers07} (reporting an exponent of $-3.8$), but
differs from the numerical calculations presented in~\cite{Ahlers07}
which obtained a power-law with an exponent of $-2$.  The difference
arises as our equation of motion includes momentum diffusion that
allows accurately accounting for the tail of $P(\dot\theta_0)$.  In
Fig.~\ref{pdftheta} we plot the experimental PDFs for $\Delta\theta\sim
\dot{\theta}$, and show that they do indeed exhibit a power-law
behavior at the tails with exponent of approximately $-4.3$ in very
good agreement with our prediction.

We thank Eric Brown and Guenter Ahlers for many helpful discussions and
for generously sharing their data with us. M.\,A. gratefully
acknowledges the Rothschild and Fulbright foundations for support.
L.\,A. is grateful for support from the Center of Excellence for
Physics of Geological Processes. This work was partially supported by
the National Science Foundation through grant number NSF-DMR-1044901.

\bibliographystyle{apsrev4-1}
\bibliography{references}

\end{document}